\documentclass[10pt]{article}
\usepackage{amsfonts}
\newcommand{\stw}{{ }}
\newcommand{\eq}{\begin{equation}}
\newcommand{\eqn}[1]{\label{#1}\end{equation}}
\newcommand{\eea}{\end{eqnarray}}
\newcommand{\eqa}{\begin{eqnarray}}
\newcommand{\eqan}[1]{\label{#1}\end{eqnarray}}
\newcommand{\ba}{\begin{array}}
\newcommand{\ea}{\end{array}}
\newcommand{\eqac}{\begin{equation}\begin{array}{rcl}}
\newcommand{\eqacn}[1]{\end{array}\label{#1}\end{equation}}

\newcommand{\ob}{{\overline{\omega}}}

\newcommand{\text}{\rm}

\begin{document}

\title{\textbf{Gribov Propagator and Symmetry Breaking:\\ a toy model}}
\author{Y. E. Chifarelli, V. E. R. Lemes
\footnote{yveseduardo@hotmail.com, vitor@dft.if.uerj.br}\\
\small\em Instituto de F\'\i sica, Universidade do Estado do Rio de Janeiro,\\
\small\em Rua S\~{a}o Francisco Xavier 524, Maracan\~{a}, Rio de
Janeiro - RJ, 20550-013, Brazil \\}
\bigskip
\maketitle

\vspace{-1cm}
\begin{abstract}

The aim of this paper is to present a connection between the
Gribov-Zwanziger condition for the mass gap and spontaneous symmetry breaking. In
order to clarify these relationship a toy model is presented and some
quantum aspects are discussed.

\end{abstract}
\setcounter{page}{0}\thispagestyle{empty}

\vfill\newpage\ \makeatother

\section{Introduction}

\hspace{.7cm}One of the most challenging issues in quantum field theory is the
understanding of the nonperturbative effects which govern the
infrared behavior of quantum field theories, this question is
particularly important in the case of Yang-Mills theory. Many
mechanisms have been proposed in an attempt to understand, at least
partially, this infrared limit. We emphasize two mechanisms in
particular, the lattice field theories \cite{Langfeld,Amemiya} and
the Gribov-Zwanziger mechanism. Recent lattice results have provided
evidences of the fact that the infrared regime of the theory is very
different from the ultraviolet one and indicates not only a mass gap
but a Gribov type propagator \cite{Attilio,Muller, Attilio2}.

The Gribov-Zwanziger framework \cite{Alkofer,Fischer,Pavlowski,Gribov,Zwanziger,BRS}
consists
in restricting the domain of integration in the Feynman path
integral within the Gribov horizon. This restriction demands the
introduction of a mass parameter and a mass gap equation and this is
a key ingredient in the Gribov mechanism \cite{Gribov,Zwanziger,Capri}.
Thus the restriction to the Gribov region is implemented
only if in addition to an action that can implement this restriction
we can also provide a gap equation to the Gribov parameter.

The achievement of the gap equation is far from easy in the case of
a general gauge choice and even more difficult to achieve if we use
the idea of location contained in the formulation of Zwanziger in a
broader sense, for example in the location of nonlocal operators in
fermionic actions. In this sense we want through a toy model to obtain
and study the Gribov-Zwanziger mechanism, in particular the setting
of the Gribov parameter without the need for conditions other than that the effective potential has a stable minimum \cite{Miransky}. Clearly the stability of the effective potential depends on the effective coupling constant is less than one. This important point will be discussed concurrently with the minimum of the potential.

The aim of this paper is to construct a toy model that presents a
gribov type propagator and an effective potential for a mass
dimension 2 condensate that can fix the Gribov mass gap due
to a non-vanishing expectation value of these condensate. It is important
to emphasize that when dealing with a toy model we can not expect all the properties present in the Gribov-Zwanziger action. For example the beta function of Yang-Mills
fields is different from that present in this toy model. It is quite clear that an action for scalar fields has no asymptotic freedom, a well known fact in the literature. However as we are dealing with a toy model, in fact as it is also made in \cite{fewr}, we can focus on common points between the model and the Gribov-Zwanziger action.

The paper
is organized as follows. In chapter 2 the toy model is presented and
the Local composite operator $(LCO)$ is introduced, also the BRST
symmetries are presented. In Chapter 3 and 4 the equations
compatible with the quantum action principle $(QAP)$ are presented
and the algebraic renormalizability of the model with the insertion
is done. In Chapter 5 the effective potential is discussed and
parameters are fixed by the renormalization group equation. Chapter
6 is dedicated to present the relation between symmetry breaking and
nonlocal BRST symmetries. Finally in the last chapter conclusions
are presented.

\section{The toy model, the LCO formalism and comparison with the Gribov-Zwanziger action}
\hspace{.7cm}Let us begin by giving the expression for a scalar $O(n)$ model that
presents the insertion of the operator
$\varphi^{a}\frac{1}{\partial^{2}}\varphi^{a}$ in a localized form
using auxiliar fields
$\chi,\overline{\chi},\omega,\overline{\omega},$ and sources $J,Q$

\begin{eqnarray}
\Sigma &=& \int d^{4}x_{E}\lbrace
\frac{1}{2}\partial_{\mu}\varphi^{a}\partial^{\mu}\varphi^{a} +
\frac{1}{2}m^{2}\varphi^{a}\varphi^{a} +
\frac{\lambda}{4!}(\varphi^{a}\varphi^{a})^{2}
+ \partial_{\mu}\overline{\chi}^{a}\partial^{\mu}\chi^{a} - \partial_{\mu}\overline{\omega}^{a}\partial^{\mu}\omega^{a} \nonumber  \\
&+& J^{ab}(\overline{\chi}^{a} - \chi^{a})\varphi^{b} +
Q^{ab}\omega^{a}\varphi^{b} - Q^{ab}(\overline{\chi}^{a} -
\chi^{a})f_{bcd}\varphi^{c}\theta^{d} + \frac{\xi}{2}J^{ab}J^{ab}
\rbrace , \label{acao}
\end{eqnarray}
the action (\ref{acao}) is left invariant under the following set of
BRST transformations:
\begin{eqnarray}
s\varphi^{a}&=& f_{abc}\varphi^{b}\theta^{c}, \,\,\, { } s\theta^{a}
= \frac{1}{2}f_{abc}\theta^{b}\theta^{c}\nonumber \\
s\overline{\omega}^{a}&=& \overline{\chi}^{a},
\,\,\,\,\,\,\,\,\,\,\,\,\,\,\, { }
s\overline{\chi}^{a} =0 \nonumber \\
s\chi^{a}&=&\omega^{a},\,\,\,\,\,\,\,\,\,\,\,\,\,\,\, { }s\omega^{a}=0 \nonumber \\
sQ^{ab}&=&J^{ab},\,\,\,\,\,\,\,\,\,\,\,\,\,\,\, { }sJ^{ab}=0. \nonumber \\
 \label{BRS}
\end{eqnarray}
Localize nonlocal terms like $\varphi^{a}\frac{1}{\partial^{2}}\varphi^{a}$ is a nontrivial task and
is done introducing the fields $\chi,\overline{\chi},\omega,\overline{\omega},$ that forms a BRST quartet,
i.e $s\overline{\omega}^{a}= \overline{\chi}^{a}$, $s\chi^{a}=\omega^{a}$ $s^{2}=0$  and the sources
$Q^{ab}$ and $J^{ab}$ in a BRST doublet structure\cite{Baulieu}. The parameter
$\xi$ has to be introduced since the introduction of the term
$J^{ab}(\overline{\chi}^{a} - \chi^{a})\varphi^{b}$ gives rise to
novel vacuum energy divergences proportional to $J^{2}$.

Here is recommended to spend a few words about the localization of nonlocal operators done with BRST quartets.
It is clear that, integrating into the fields $\overline{\chi}$ and $\chi$ we have
\begin{equation}
\partial_{\mu}\overline{\chi}^{a}\partial^{\mu}\chi^{a}
+ J^{ab}(\overline{\chi}^{a} - \chi^{a})\varphi^{b} \Longrightarrow
- J^{ab}\varphi^{b}\frac{1}{\partial^{2}}J^{ac}\varphi^{c}.
\end{equation}
If we find one equation that gave us a nonzero value for the souce
$J^{ab}$ the nonlocal operator changes the original klein-Gordon
propagator to a Gribov type one\cite{Gribov}. This task is one of
the most important in the Gribov procedure. Unfortunately is not
possible to obtain, from geometrical considerations, a gap equation
in any case,for example in our present case in which there are no
null eigenvalues associated to the equation $(-\partial^{2}+
m^{2})\varphi = \varrho \varphi$. In more complex actions like
Yang-Mills in the Feynmann gauge, for example, it is also impossible
to obtain a gap equation from geometrical considerations and a gap
equation is imposed ``by hand``. In these sense, the $LCO$ procedure
together with the localization process appears to be a natural way
to obtain nonzero values for $J$. For a detailed introduction to the
local composite operator (LCO) formalism and to the algebraic
renormalization technique, the reader is referred to
\cite{LCO1,LCO2}, respectively. Again it is important to emphasize
that the scalar model is a toy model and we are using this model to
study the procedure of symmetry breaking that can be also applied to
the Gribov-Zwanziger action. This procedure is most easily
understood with the toy model as a guide, of course not all
properties of the Gribov-Zwanziger action are presented in the toy
model. For example, in the strong coupling limit of Yang-Mills, the
effective coupling constant present in the calculation of the
minimun of  the potential in Gribov-Zwanziger action is less than
one. This is not the case  of the scalar field action. We will not
deal with the renormalizability of the Gribov-Zwanziger action since
it was done into many references, as for example in
\cite{Gribov,Zwanziger,Schaden,Zwan}. However, it is convenient to
present the similarities and differences between this action and the
toy model presented. For detailed calculations of the
renormalizability of the action of Gribov-Zwanziger in the Landau
gauge follow \cite{Schaden}. In order to constrain the gauge fields
to the Gribov region and have a local action in the fields, a BRST
quartet structure $(\omega^a_i, \varphi^a_i, \bar\omega^{ai},
\bar\varphi^{ai})$ and a pair of BRST sources are introduced
$(\overline{Q}^{ai}_\mu,Q^{a}_{\mu i}, \overline{J}^{ai}_\mu,
J^{a}_{\mu i})$. The BRST for the fields and sources are:
\begin{eqnarray}
s A^{a}_{\mu} &=& - {(D_{\mu}c)}^{a} \nonumber \\
s c^{a} &=& {1 \over 2}f^{abc}c^b c^c \nonumber \\
s {\overline{c}}^{a}&=& ib^{a}, \,\,\,\,\, s b^{a}=0 \nonumber \\
s{\bar \omega}^{ai} &=& {\overline{\varphi}}^{ai}, \,\,\,\,\,  s{\overline{\varphi}}^{ai}=0\nonumber \\
s\varphi^a_i &=& \omega^a_i , \,\,\,\,\, s\omega^a_i=0 \nonumber \\
s \overline{Q}^{ai}_\mu &=& \overline{J}^{ai}_\mu , \,\,\,\,\, s \overline{J}^{ai}_\mu=0\nonumber \\
s Q^a_{\mu i} &=& J^a_{\mu i} , \,\,\,\,\, s J^a_{\mu i}=0 \nonumber \\
\label{brsgz}
\end{eqnarray}
These new
fields transform under a global $U(f)$ symmetry on the composite
index~\mbox{$i=(\nu, b)$},  with $f=4(N^2-1)$.
The localized, BRST invariant version of the Gribov-Zwanziger action is given by:
\eq\ba{rl}
S =\int d^{4}x_{E}\lbrace & \frac{1}{4}F^{a}_{\mu\nu}F^a_{\mu\nu}
-(i\partial_\mu b^a) A^a_\mu -(\partial_\mu{\overline{c}}^a) (D_\mu c)^a\\
&
+(\partial_\mu\overline{\varphi}^{ai}) (D_\mu \varphi_i)^a
-(\partial_\mu\overline{\omega}^{ai}) (D_\mu \omega_i)^a
+f^{abc}(\partial_\mu \ob^{ai})(D_\mu c)^b \varphi_i^c\\
&
+ \overline{J}^{ai}_\mu gf_{abc}A^b_\mu {\varphi}^{ci}
- \overline{Q}^{ai}_\mu gf_{abc}A^b_\mu {\omega}^{ci}
+ gf^{abc}\overline{Q}^{ai}_\mu (D_\mu c)^b \varphi_i^c\\
&+ J^a_{\mu i} gf_{abc}A^b_\mu \overline{\varphi}^{ci}
+ Q^a_{\mu i} gf_{abc}A^b_\mu \overline{\omega}^{ci}
-gf^{abc}J^a_{\mu i}(D_\mu c)^b\ob^{ci}
+ \xi\overline{J}^{ai}_\mu J^a_{\mu i}
- \xi\overline{Q}^{ai}_\mu Q^a_{\mu i} \\
&+ \Omega^a_\mu (sA^a_\mu) + L^a (sc^a)\rbrace \ ,
\ea\eqn{acao-fontes} where we can see that the sources
$\overline{J}^{ai}_\mu$ and $J^a_{\mu i}$ are coupled to
$gf^{abc}{\overline{\varphi}}^{bi}A^{c}_{\mu},gf^{abc}{\varphi}^{bi}A^{c}_{\mu}$,
defining a local composite operator of UV dimension $2$, like in the
toy model. As a demonstration of the renormalizability of the
Gribov-Zwanziger action is extensive and has been held in many
gauges, we will pass to the analyses required for the algebraic
renormalizability of the toy model.

\section{Equations compatible with the Quantum Action Principle}

\hspace{.7cm}Now we will present several symmetries of the action
$\Sigma$ that are compatible with the Quantum Action Principle
(QAP)\cite{livro}, which will be useful in the BRST renormalization
procedure. First we have equations of motion with classical
breaking.

\begin{eqnarray}
\frac{\delta\Sigma}{\delta\overline{\omega}^{a}} &=& \partial^{2}\omega^{a},\,\,\,\,\,
\frac{\delta\Sigma}{\delta\omega^{a}} = -\partial^{2}\overline{\omega}^{a} + Q^{ab}\varphi^{b},\nonumber  \\
\frac{\delta\Sigma}{\delta\overline{\chi}^{a}} &=& -\partial^{2}\chi^{a} + J^{ab}\varphi^{b} - Q^{ab}f_{bcd}\varphi^{c}\theta^{d}, \nonumber  \\
\frac{\delta\Sigma}{\delta\chi^{a}} &=&
-\partial^{2}\overline{\chi}^{a} - J^{ab}\varphi^{b} +
Q^{ab}f_{bcd}\varphi^{c}\theta^{d}.
 \label{classicalbreak}
\end{eqnarray}
It is important to emphasize that the ghost fields $\theta^{a}$ are
global and therefore classical fields as well as the sources.

The Slavnov-Taylor Identity:

\begin{equation}
S(\Sigma)=\int d^{4}x_{E}\lbrace
f_{abc}\varphi^{b}\theta^{c}\frac{\delta\Sigma}{\delta \varphi^{a}}
+ \overline{\chi}^{a}\frac{\delta\Sigma}{\delta
\overline{\omega}^{a}} + \omega^{a}\frac{\delta\Sigma}{\delta
\chi^{a}} + J^{ab}\frac{\delta\Sigma}{\delta Q^{ab}}\rbrace +
\frac{1}{2}f_{abc}\theta^{b}\theta^{c}\frac{\delta\Sigma}{\delta
\theta^{a}} =0,
 \label{Slavnov}
\end{equation}
a global ghost equation
\begin{eqnarray}
\mathcal{G}^{a}(\Sigma )&=& \Delta^{a}\nonumber \\
\mathcal{G}^{a}&=& \frac{\delta}{\delta\theta^{a}} - \int d^{4}x_{E}
( f_{abc} Q^{ib}\frac{\delta}{\delta J^{ic}} ),\,\,\,\,\,\,\,
\Delta^{a} = -\int d^{4}x_{E}\,\,\xi \,\, f_{abc} Q^{ib}J^{ic},
\label{ghostglobal}
\end{eqnarray}
the rigid symmetry
\begin{equation}
\mathcal{W}^{a}(\Sigma ) = \int d^{4}x_{E}( f_{abc}\lbrace
\varphi^{b}\frac{\delta\Sigma}{\delta \varphi^{c}} +
J^{ic}\frac{\delta\Sigma}{\delta J^{ib}} +
Q^{ic}\frac{\delta\Sigma}{\delta Q^{ib}} \rbrace )+
f_{abc}\theta^{b}\frac{\delta\Sigma}{\delta \theta^{c}} =0,
 \label{group}
\end{equation}
and a linearly broken symmetry which involves the ghosts of the
quartet, the source $Q^{ab}$ and the global ghost $\theta^{a}$
\begin{eqnarray}
\mathcal{Q}(\Sigma)&=& \Pi \nonumber \\
\mathcal{Q} &=& \int d^{4}x_{E}(
\overline{\omega}^{a}\frac{\delta}{\delta\overline{\omega}^{a}} -
\omega^{a}\frac{\delta}{\delta\omega^{a}} +
Q^{ab}\frac{\delta}{\delta Q^{ab}} +
f_{bcd}Q^{ab}\theta^{d}\frac{\delta}{\delta J^{ab}}) \nonumber \\
\Pi &=& \int d^{4}x_{E}\,\,\xi\,\, f_{bcd}Q^{ab}J^{ac}\theta^{d}.
\label{reality}
\end{eqnarray}
These equations, together, provide us all the constraints of the
classical action $\Sigma$ that can be extended to the quantum
action.

\section{Stability of the quantum action}

\hspace{.7cm}In order to study the stability of the quantum action let us start
by presenting the quantum numbers of all fields and sources:
\begin{table}[h]
\centering
\begin{tabular}{|c|c|c|c|c|c|c|c|c|}
\hline fields and sources & $\stw{\varphi}$ & $\stw{\theta}$ &
$\stw{\overline{\chi}}$ & $\stw{\chi}$ & $\stw{\overline{\omega}}$ &
$\stw{\omega}$ &  $\stw{J}$ &
 $\stw{Q}$ \\
\hline
UV dimension & 1 & 0 & 1 & 1 & 1 & 1 & 2 & 2 \\
Ghost number & 0 & 1 & 0 & 0 & $-1$ & 1 & 0 & $-1$ \\
Statistics & co & an & co & co & an & an & co & an \\
\hline
\end{tabular}
\caption{Quantum numbers of fields and sources.} \label{table1}
\end{table}
We remark to the fact that in the stability analysis of the quantum
action it is necessary to take into account that the ghost $\theta $
is a global ghost and only characterizes the rotation symmetry.

\subsection{The invariant counterterm}

In order to characterize any invariant counterterm which can be
added freely to all orders in perturbation theory \cite{livro}, we
perturb the classical action $\Sigma$ by adding an arbitrary
integrated local polynomial $\Sigma^{count}$ of dimension up-bounded
by four, vanishing ghost number and obeying all the other symmetries
that are linearly broken. We demand that
$\Gamma=\Sigma+\epsilon\Sigma^{count}+O(\epsilon^2)$, where
$\epsilon$ is a small expansion parameter, satisfies the same Ward
identities as $\Sigma$. This requirement provides the following
constraints on the counterterm:

\begin{eqnarray}
\frac{\delta\Sigma^{count}}{\delta\overline{\omega}^{a}} &=& 0,\label{1}  \\
\frac{\delta\Sigma^{count}}{\delta\omega^{a}} &=& 0,\label{2}  \\
\frac{\delta\Sigma^{count}}{\delta\overline{\chi}^{a}} &=& 0, \label{3}  \\
\frac{\delta\Sigma^{count}}{\delta\chi^{a}} &=& 0, \label{4}\\
{\cal B}_{\Sigma}\Sigma^{count}&=&0, \label{5} \\
\mathcal{G}^{a}(\Sigma^{count} )&=& 0, \label{6} \\
\mathcal{W}^{a}(\Sigma^{count} )&=& 0, \label{7} \\
\mathcal{Q}(\Sigma^{count})&=&0, \label{8}
\end{eqnarray}
where in (\ref{5}), ${\cal B}_{\Sigma}$ stands for the nilpotent
Slavnov-Taylor operator,
\begin{equation}
{\cal B}_{\Sigma}=\int d^{4}x_{E}\lbrace
f_{abc}\varphi^{b}\theta^{c}\frac{\delta}{\delta \varphi^{a}} +
\overline{\chi}^{a}\frac{\delta}{\delta \overline{\omega}^{a}} +
\omega^{a}\frac{\delta}{\delta \chi^{a}} +
J^{ab}\frac{\delta}{\delta Q^{ab}}\rbrace +
\frac{1}{2}f_{abc}\theta^{b}\theta^{c}\frac{\delta}{\delta
\theta^{a}} =0.
 \label{Slavnov}
\end{equation}
The set of equations (\ref{1},\ref{2},\ref{3},\ref{4}) are of
particular importance in order to obtain the counterterm action. Due
to the fact that they are local equations on the fields of the
quartet. Thus the fields
$\overline{\omega},\omega,\overline{\chi},\chi$ do not appear in
$\Sigma^{count}$.
Equations (\ref{5},\ref{6},\ref{7},\ref{8}) are responsible for
reducing even more the set of fields and sources from which
$\Sigma^{count}$ may depend.
At the end, the counterterm action is only a functional of the field
$\varphi^{a}$ of the form:

\begin{equation}
\Sigma^{count}= \int d^{4}x_{E} \lbrace
\frac{a_{1}}{2}\partial_{\mu}\varphi^{a}\partial_{\mu}\varphi^{a} +
\frac{a_{2}}{2}m^{2}\varphi^{a}\varphi^{a} +
\frac{a_{3}}{4!}\lambda(\varphi^{a}\varphi^{a})^{2} \rbrace .
\label{contaction}
\end{equation}
This result is extremely important in the analisys of the quantum
behavior of such model. In fact it indicates that all the fields
used in the localization process
$\overline{\omega},\omega,\overline{\chi},\chi$, the
sources$\stw{J}$ and $\stw{Q}$ and the parameter $\xi$ do not
renormalizes independently.
It is immediate to check that the counterterm action can be reabsorbed into the classical action $\Sigma$
\begin{equation}
\Sigma + \varepsilon \Sigma^{count} =
\Sigma(\lambda_{0},\xi_{0},m_{0},\varphi^{a}_{0},\overline{\omega}^{a}_{0},\omega^{a}_{0},\overline{\chi}^{a}_{0},\chi^{a}_{0}) + O(\varepsilon^{2}),
\end{equation}
by redefining mass, sources, couplings and field amplitudes according to
\begin{eqnarray}
\lambda_{0} &=& Z_{\lambda} \lambda \hspace{1.3 cm} \xi_{0} = Z_{\xi}\xi \hspace{1cm} m^{2}_{0} = Z_{m}m^{2} \nonumber \\
\varphi^{a}_{0} &=& Z^{\frac{1}{2}}_{\varphi}\varphi^{a}\nonumber \\
\overline{\omega}^{a}_{0}
&=& Z^{\frac{1}{2}}_{\overline{\omega}}\overline{\omega}^{a} \hspace{1cm}\omega^{a}_{0} = Z^{\frac{1}{2}}_{\omega}\omega^{a} \nonumber \\
\overline{\chi}^{a}_{0} &=& Z^{\frac{1}{2}}_{\overline{\chi}}\overline{\chi}^{a} \hspace{1cm} \chi^{a}_{0} = Z^{\frac{1}{2}}_{\chi}\chi^{a} \nonumber \\
J^{ab}_{0} &=& Z_{J} J^{ab} \hspace{.7cm} Q^{ab}_{0} = Z_{Q} Q^{ab} \nonumber \\
\label{defZ}
\end{eqnarray}
with
\begin{eqnarray}
Z_{\lambda} &=& 1 + \varepsilon \textit{z}_{\lambda} = 1 + \varepsilon (a_{3} - 2a_{1}) \nonumber \\
Z_{\xi} &=& 1 + \varepsilon\textit{z}_{\xi} = 1 + \varepsilon a_{1}\nonumber \\
Z_{m} &=& 1 + \varepsilon\textit{z}_{m} = 1 + \varepsilon (a_{2} - a_{1})\nonumber \\
Z^{\frac{1}{2}}_{\varphi} &=& 1 + \frac{\varepsilon}{2}\textit{z}_{\varphi} = 1 + \frac{\varepsilon}{2}a_{1}\nonumber \\
Z^{\frac{1}{2}}_{\overline{\omega}}&=& 1 + \frac{\varepsilon}{2}\textit{z}_{\overline{\omega}} = 1 \hspace{1cm} Z^{\frac{1}{2}}_{\omega} = 1 + \frac{\varepsilon}{2} \textit{z}_{\omega} = 1\nonumber \\
Z^{\frac{1}{2}}_{\overline{\chi}}&=& 1 + \frac{\varepsilon}{2}\textit{z}_{\overline{\chi}} = 1 \hspace{1cm} Z^{\frac{1}{2}}_{\chi} = 1 + \frac{\varepsilon}{2} \textit{z}_{\chi} = 1\nonumber \\
Z_{J}= Z_{Q} &=& 1 + \varepsilon\textit{Z}_{Q} = 1 -
\frac{\varepsilon}{2}a_{1}. \label{defzp}
\end{eqnarray}
In possession of the renormalization relations, we can now turn to
the calculation of the effective potential.

\section{Effective potential at one loop in $\Sigma$ and similarities with the Gribov-Zwanziger case}

\hspace{.7cm}The first step in order to study the condensate $(\overline{\chi}^{a} - \chi^{a})\varphi^{b}$ is to analyse
the generating functional \textit{W}(j). Setting thus to zero the external source
$Q^{ab}$ we have

\begin{equation}
exp(-\textit{W}(j))= \int [D\phi]exp-\lbrace S_{O(n)} + \int d^{4}x_{E} (J^{ab}(\overline{\chi}^{a} - \chi^{a})\varphi^{b} + \frac{\xi}{2}J^{ab}J^{ab})\rbrace,
\label{wj}
\end{equation}
where $[D\phi]$ denotes integration over all quantum fields, i.e
$\varphi,\chi,\overline{\chi},\omega,\overline{\omega}$ and

\begin{equation}
S_{O(n)} = \int d^{4}x_{E}
\{\frac{1}{2}\partial_{\mu}\varphi^{a}\partial_{\mu}\varphi^{a} +
\frac{m^{2}}{2}\varphi^{a}\varphi^{a} +
\frac{\lambda}{4!}(\varphi^{a}\varphi^{a})^{2} +
\partial_{\mu}\overline{\chi}^{a}\partial_{\mu}\chi^{a} - \partial_{\mu}\overline{\omega}^{a}\partial_{\mu}\omega^{a}\}. \label{acaoOn}
\end{equation}
Taking the functional derivative of the above expression we obtain

\begin{equation}
\frac{\delta\textit{W}(j)}{\delta J^{ab}}|_{J^{ab}=0}=
\langle(\overline{\chi}^{a} -
\chi^{a})\varphi^{b}\rangle.\label{j-cond}
\end{equation}

Before introducing the Hubbard-Stratonovich field is interesting to note that the following combination
of sources and fields has the property of renormalise as exactly the field $\varphi$

\begin{eqnarray}
\mu^{ab} &=& \xi J^{ab} + (\overline{\chi}^{a}-\chi^{a})\varphi^{b},\nonumber \\
\mu^{ab}_{0} &=& Z_{\varphi}Z^{-\frac{1}{2}}_{\varphi}\xi J^{ab} +
Z^{\frac{1}{2}}_{\varphi}(\overline{\chi}^{a}-\chi^{a})\varphi^{b} \nonumber \\
\mu^{ab}_{0} &=& Z^{\frac{1}{2}}_{\varphi}\mu^{ab}.
\label{redeffield}
\end{eqnarray}
In order to deal with the term $\frac{\xi}{2}J^{ab}J^{ab}$ we follow \cite{knecht}, introducing a Hubbard-Stratonovich field $\sigma^{ab}$ so that

\begin{equation}
\frac{\xi}{2}J^{ab}J^{ab} +
J^{ab}(\overline{\chi}^{a}-\chi^{a})\varphi^{b} =
\frac{1}{2\xi}(\mu^{ab}\mu^{ab} -
(\overline{\chi}^{a}-\chi^{a})\varphi^{b}(\overline{\chi}^{a}-\chi^{a})\varphi^{b})
\end{equation}

\begin{equation}
\frac{1}{2\xi}\mu^{ab}\mu^{ab} =
-\frac{1}{2\xi}\sigma^{ab}\sigma^{ab} +
\frac{1}{\xi}\sigma^{ab}\lbrace \xi J^{ab}+(\overline{\chi}^{a}-\chi^{a})\varphi^{b}\rbrace.
\end{equation}
As a consequence, for the functional generator, we get
\begin{equation}
exp(-\textit{W}(j))= \int [D\phi]exp-\lbrace S(\varphi,\sigma) + \int d^{4}x_{E} \sigma^{ab}J^{ab}\rbrace,
\label{wj}
\end{equation}
with
\begin{equation}
S(\varphi,\sigma) = S_{O(n)} + \int d^{4}x_{E}\lbrace
-\frac{1}{2\xi}\sigma^{ab}\sigma^{ab} +
\frac{1}{\xi}\sigma^{ab}(\overline{\chi}^{a}-\chi^{a})\varphi^{b}
-\frac{1}{2\xi}(\overline{\chi}^{a}-\chi^{a})\varphi^{b}(\overline{\chi}^{a}-\chi^{a})\varphi^{b}\rbrace,
\end{equation}
\begin{equation}
\frac{\delta\textit{W}(j)}{\delta J^{ab}}|_{J^{ab}=0}= \langle(\overline{\chi}^{a} - \chi^{a})\varphi^{b}\rangle = \langle \sigma^{ab} \rangle.
\end{equation}
This identity says that the condensate $\langle(\overline{\chi}^{a}
- \chi^{a})\varphi^{b}\rangle$ is related to the nonvanishing value
of $\sigma^{ab}$ calculated with the action $S(\varphi,\sigma)$. It
is important to emphasize that due to the intrinsic  non
perturbative caracter of the LCO approach it is not necessary to go
beyond the one loop level calculations to see the nonperturbative
features. Higher loop calculations give us only better numerical
results for the value of the condensate, as this is a toy model,
numerical refinements are not needed and do not affect the
understanding of the proposed method.Again it is important to
emphasize here that we know that a scalar model is not
asymptotically free. However the Gribov-Zwanziger action has
nonabelian gauge fields which have the property of being
asymptotically free, as we know from the sign of the beta function.
We believe, however, that the toy model can be used to more easily
understand the mechanism present in the Gribov-Zwanziger action.
Similarly in the Gribov-Zwanziger action we have to do the same kind
of analysis in order to calculate the effective potential, $i.e.$
setting to zero the sources $Q^{a}_{\mu \hspace{1mm}i}$ and
$\overline{Q}^{a}_{\mu \hspace{1mm}i}$ and repeat the same procedure
performed in the toy model.

\begin{eqnarray}
exp(-\textit{W}_{GZ}(j))&=& \int [D\phi]exp-\lbrace S_{A,\overline{\varphi},\varphi,\overline{\omega},\omega} + S_{\overline{J},J}\rbrace \nonumber \\
S_{\overline{J},J} &=&
\int d^{4}x_{E} (\overline{J}^{ai}_\mu gf_{abc}A^b_\mu {\varphi}^{ci}
+ J^a_{\mu i} gf_{abc}A^b_\mu \overline{\varphi}^{ci} -gf^{abc}J^a_{\mu i}(D_\mu c)^b\ob^{ci}
+ \xi\overline{J}^{ai}_\mu J^a_{\mu i}) \nonumber \\
S_{A,\overline{\varphi},\varphi,\overline{\omega},\omega} &=& \int d^{4}x_{E}\lbrace \frac{1}{4}F^{a}_{\mu\nu}F^a_{\mu\nu}
-(i\partial_\mu b^a) A^a_\mu -(\partial_\mu{\overline{c}}^a) (D_\mu c)^a \nonumber \\
&+&(\partial_\mu\overline{\varphi}^{ai}) (D_\mu \varphi_i)^a
-(\partial_\mu\overline{\omega}^{ai}) (D_\mu \omega_i)^a
+f^{abc}(\partial_\mu \ob^{ai})(D_\mu c)^b \varphi_i^c \rbrace \ ,
\label{wjgz}
\end{eqnarray}

The functional derivative with respect to the sources $\overline{J}^{ai}_\mu$ and $J^{ai}_\mu$ gave us
\begin{eqnarray}
\frac{\delta\textit{W}_{GZ}(j)}{\delta J^{ai}_{\mu}}|_{\overline{J}^{ai}_{\mu}=0}&=&
gf_{abc}\langle A^b_\mu {\varphi}^{ci}\rangle - gf^{abc}\langle (D_\mu c)^b\ob^{ci}\rangle.\nonumber \\
\frac{\delta\textit{W}_{GZ}(j)}{\delta \overline{J}^{ai}_{\mu}}|_{J^{ai}_{\mu}=0}&=&
gf_{abc}\langle A^b_\mu \overline{\varphi}^{ci}\rangle ,
\end{eqnarray}
and, like in the toy model, we can introduce the adequate
Hubbard-Stratonovich fields and rewrite the
$\textit{W}_{GZ}(\sigma)$ as:
\begin{equation}
exp(-\textit{W}_{GZ}(\sigma))= \int [D\phi]exp-\lbrace
S_{GZ}(\sigma) + \int d^{4}x_{E} (\overline{\sigma}^{ai}_\mu
{J}^{ai}_\mu + {\sigma}^{ai}_\mu \overline{J}^{ai}_\mu ) \rbrace,
\label{wjGZ}
\end{equation}
with
\begin{eqnarray}
S_{GZ}(\sigma) &=&
S_{A,\overline{\varphi},\varphi,\overline{\omega},\omega} \nonumber
\\ &+& \int d^{4}x_{E} \{-\frac{1}{\xi}\overline{\sigma}^{ai}_\mu \sigma^{ai}_\mu +
\frac{1}{\xi}{\sigma}^{ai}_\mu [ gf_{abc}( A^b_\mu
\overline{\varphi}^{ci})] + \frac{1}{\xi}\overline{\sigma}^{ai}_\mu[
 gf_{abc}( A^b_\mu {\varphi}^{ci} - gf^{abc}
(D_\mu c)^b\ob^{ci})]\} \nonumber \\ &-&\int d^{4}x_{E} \{
\frac{1}{\xi}g^{2}f_{abc}A^b_\mu {\varphi}^{ci}f_{ade}A^d_\mu
\overline{\varphi}^{ei} + \frac{1}{\xi}g^{2}f_{abc}A^b_\mu
{\varphi}^{ci}f_{ade}(D_\mu c)^d\ob^{ei} \} ,
\end{eqnarray}
\begin{equation}
\frac{\delta\textit{W}_{GZ}(j)}{\delta
J^{ai}_{\mu}}|_{\overline{J}^{ai}_{\mu}=0}= \langle{\sigma}^{ai}_\mu
\rangle \hspace{3cm} \frac{\delta\textit{W}_{GZ}(j)}{\delta
\overline{J}^{ai}_{\mu}}|_{J^{ai}_{\mu}=0}= \langle
\overline{\sigma}^{ai}_\mu \rangle .
\end{equation}
Clearly the difference between the calculation with Gribov-Zwanziger
and the toy model is the coupling constant. In the case of
Gribov-Zwanziger, the coupling with $\frac{g^{2}}{\xi}$ indicates
that $\xi $ is proportional to $\frac{1}{g^{2}}$. Thus, in the
infrared limit the effective potential, which is proportional to
$\xi$ allows perturbative calculations and present a well-defined
mass gap, indeed the same result is also obtained by the gap
equation in Gribov-Zwanziger.

\subsection{Evaluation of the toy model effective potential at one loop}

\hspace{.7cm}In order to compute the effective potential for $\sigma^{ab}$ at the one-loop order only
the quadratic part of the action $S(\varphi,\sigma)$ is relevant, namely

\begin{equation}
S_{quad}= \int d^{4}x_{E} \lbrace
\frac{\sigma_{ab}\sigma_{ab}}{2\xi} +
\phi_{a}^{\dagger}\mathcal{M}_{ab}\phi_{b} \rbrace
\end{equation}
where $\phi_{a}^{\dagger}= (\varphi^{a},\overline{\chi}^{a})$, and
\begin{equation}
\mathcal{M}^{ab}=\left(
\begin{tabular}{ll}
$\frac{1}{2}(k^{2}+m^{2})\delta_{ab}$ & $-\frac{1}{\xi}\sigma_{ab}$ \\
$\frac{1}{\xi}\sigma_{ab}$ & $k^{2}\delta_{ab}$%
\end{tabular}
\right) \;.  \label{mab}
\end{equation}

After straight forward calculations, using dimensional regularization and in the $\overline{MS}$ scheme the one loop effective potential is giben by:
\begin{equation}
V_{eff}= \frac{\sigma_{ab}\sigma_{ab}}{2\xi} +
\hbar^{3}\frac{n(n-1)}{128\pi^{2}}(m^{4}-2\frac{\sigma_{ab}\sigma_{ab}}{\xi^{2}(n)})
\lbrace\ln(\frac{2\frac{\sigma_{ab}\sigma_{ab}}{\xi^{2}(n)}}{16\pi^{2}\overline{\Lambda}^{4}})
-\frac{5}{6}\rbrace,
\end{equation}
whose minimum of the effective potential is given by the condition
\begin{equation}
\sigma^{ab}_{min}\sigma^{ab}_{min} =
4\pi^{2}\xi^{2}n(n-1)\overline{\Lambda}^{4}exp(\frac{16\pi^{2}\xi}{\hbar^{3}}).
\end{equation}

According now to \cite{knecht}, the parameter $\xi$ can be computed order by order in
the loop expansion $\xi= \xi_{0} + \hbar\xi_{1} + \hbar^{2}\xi_{2} + ...$ and is obtained from
the renormalization group equations.This requirement enables the LCO technique to see
nonperturbative effects. In fact \cite{knecht} shows that this coefficient is scheme independent
and plays a crucial hole in order to obtain a nontrivial vacuum configuration
for $\langle(\overline{\chi}^{a} - \chi^{a})\varphi^{b}\rangle$.
Let us now proceed with the evaluation of the parameter $\xi$ at first order in $\hbar$
requiring that:

\begin{equation}
\Lambda\frac{d V_{eff}(\sigma)}{d \Lambda}=0 + O(\hbar).
\end{equation}
Based on the requirement above and following the references\cite{knecht, Lemes,Dudal}, after some standard calculations for the
method to obtain the LCO parameter $\xi$

\begin{equation}
\xi_{0}= \frac{n(n-1)}{64\pi^{2}\gamma_{\varphi^{(1)}}},
\end{equation}

\begin{equation}
\gamma_{\varphi^{(1)}} =
\frac{(n+2)}{3}\frac{\lambda^{2}}{12(4\pi)^{4}}
\end{equation}
completing therefore the evaluation of the one-loop effective
potential for $(\overline{\chi}^{a}-\chi^{a})\varphi^{b}$. Important
remarks are now in order to be presented.

\begin{itemize}
\item By combining the LCO technique with the BRST algebraic renormalization
we have been able to obtain the one-loop effective potential for
$(\chi^{a}-\chi^{a})\varphi^{b}$.It is clear that in the toy model
this potential do not correspond to a stable mass gap due to the
differences between the scalar interaction and the Gribov-Zwanziger
interaction. However by construction, the effective potential
$V_{eff}(\sigma)$ obeys the renormalization group equation turning
possible to determine that the vacuum non zero expectation value is
stable in the Gribov-Zwanziger case. It is also clear that, in the Gribov-Zwanziger case the parameter $\xi_{0}$ is proportional to the inverse of $g^{2}$ i.e. $\xi_{0} \propto \frac{1}{g^{2}}$.

\item The propagator for $\varphi$, in the condensed vacuum, becames now a Gribov type propagator and is given by:
\begin{equation}
<\varphi^{a}(k)\varphi^{b}(-k)> = \frac{k^{2}}{k^{4} + m^{2}k^{2} +
\frac{v^{4}}{\xi^{2}}},
\end{equation}
where
\begin{equation}
\sigma^{ab}= v^{2}\delta^{ab}.
\end{equation}
This result indicates that in an action with BRST quartets and an
insertion of $UV$ dimension 4 a Gribov type propagator will appear,
if there is a stable condensed vacuum. We emphasize again that the
LCO mechanism allows us to determine when a Gribov type propagator
is favored or not simply analizing the vacuum expectation of the
effective potential.

\item Due to the characteristics of the $LCO$ method, exactly the same results of the usual gap equation can be obtained in the
case of the Gribov-Zwanziger action. The method however is
applicable to other actions in which the geometrical motivation,
given by restriction to the Gribov region, for the gap equation can
not be obtained in a direct way. An example of a Yang-Mills action
presenting a Gribov type propagator, whose equation of gap should be
placed as an extra condition without direct geometric motivation is
given by \cite{sps}.
\end{itemize}

\section{How symmetry breaking changes the BRST}

\hspace{.7cm}Since we have a non zero expectation value for
$<(\overline{\chi}^{a}-\chi^{a})\varphi^{b}>$ and due to equation
(\ref{j-cond}),which relates the value of these condensate to the
source $J^{ab}$, it is natural to redefine these source as $J^{ab} =
\widetilde{J}^{ab} + \frac{v^{2}}{\xi}\delta^{ab}$, where
$\widetilde{J}^{ab}$ has zero expectation value. Substituting
$\widetilde{J}^{ab} + \frac{v^{2}}{\xi}\delta^{ab}$ in the Slavnov
Taylor identity we obtain for the action $\Sigma_{v\neq 0}$, which
is defined by:
\begin{equation}
 \Sigma_{v\neq 0}= \Sigma_{(J+\frac{v^{2}}{\xi})}|_{J=0,Q=0},
\end{equation}
the following Slavnov Taylor relation
\begin{eqnarray}
S(\Sigma_{v\neq 0})&=&\int d^{4}x_{E}\lbrace
f_{abc}\varphi^{b}\theta^{c}\frac{\delta\Sigma_{v\neq 0}}{\delta
\varphi^{a}} + \overline{\chi}^{a}\frac{\delta\Sigma_{v\neq
0}}{\delta \overline{\omega}^{a}} +
\omega^{a}\frac{\delta\Sigma_{v\neq 0}}{\delta \chi^{a}} +
(\widetilde{J}^{ab} +
\frac{v^{2}}{\xi}\delta^{ab})\frac{\delta\Sigma_{v\neq 0}}{\delta
Q^{ab}}\rbrace \nonumber \\ &+&
\frac{1}{2}f_{abc}\theta^{b}\theta^{c}\frac{\delta\Sigma_{v\neq
0}}{\delta
\theta^{a}} \nonumber \\
&=&\int d^{4}x_{E}\lbrace -\frac{v^{2}}{\xi}\omega^{a}\varphi^{a} -
\frac{v^{2}}{\xi}(\overline{\chi}^{a} -
{\chi}^{a})f_{abc}\theta^{b}\varphi^{c} \rbrace ,
 \label{n-Slavnov}
\end{eqnarray}
showing that $\Sigma_{v\neq 0}$ is not invariant under the original BRST
symmetry and the breaking term has ultraviolet dimension $2$ which
characterizes a soft breaking term.

Another equation compatible with the quantum action principle is
given by
\begin{equation}
\frac{\delta\Sigma_{v\neq 0}}{\delta\overline{\omega}^{a}} =
\partial^{2}\omega^{a}.
\end{equation}
This equation can be solved for
\begin{equation}
\omega^{a}(x)=-(\frac{1}{-\partial^{2}})\frac{\delta\Sigma_{v\neq
0}}{\delta\overline{\omega}^{a}} \equiv -\int d^{4}y_{E}\lbrace
(\frac{1}{-\partial^{2}})_{xy}\frac{\delta\Sigma_{v\neq
0}}{\delta\overline{\omega}^{a}(y)}\rbrace .
\end{equation}

The term $-\frac{v^{2}}{\xi}(\overline{\chi}^{a} -
{\chi}^{a})f_{abc}\theta^{b}\varphi^{c}$ also can be solved with the
use of 2 equations compatible with the quantum action principle. The
equations are:
\begin{eqnarray}
\frac{\delta \Sigma_{v\neq 0}}{\delta \overline{\chi}^{a}} &=&
-\partial^{2}{\chi}^{a} + \frac{v^{2}}{\xi}\varphi^{a} \nonumber \\
\frac{\delta \Sigma_{v\neq 0}}{\delta {\chi}^{a}} &=&
-\partial^{2}\overline{\chi}^{a} - \frac{v^{2}}{\xi}\varphi^{a}.
\end{eqnarray}

Therefore the breaking term of (\ref{n-Slavnov}) can be rewritten as
\begin{equation}
S(\Sigma_{v\neq 0}) - \int d^{4}x_{E}\frac{v^{2}}{\xi}\lbrace
(\frac{1}{-\partial^{2}})\varphi^{a}\frac{\delta\Sigma_{v\neq
0}}{\delta\overline{\omega}^{a}}  -
f_{abc}\theta^{b}(\frac{1}{-\partial^{2}})\varphi^{c}(\frac{\delta
\Sigma_{v\neq 0}}{\delta {\chi}^{a}} - \frac{\delta \Sigma_{v\neq
0}}{\delta \overline{\chi}^{a}}\rbrace = 0,
\end{equation}
and the action $\Sigma_{v\neq 0}$ is left invariant under the new
set of BRST transformations:

\begin{eqnarray}
s'\varphi^{a}&=& f_{abc}\varphi^{b}\theta^{c} \nonumber \\
s'\theta^{a}&=& \frac{1}{2}f_{abc}\theta^{b}\theta^{c} \nonumber \\
s'\overline{\omega}^{a}&=& \overline{\chi}^{a} -
\frac{v^{2}}{\xi}(\frac{1}{-\partial^{2}})\varphi^{a} \nonumber \\
s'\overline{\chi}^{a} &=& -\frac{v^{2}}{\xi}f_{abc}\theta^{b}(\frac{1}{-\partial^{2}})\varphi^{c} \nonumber \\
s'\chi^{a}&=&\omega^{a} + \frac{v^{2}}{\xi}f_{abc}\theta^{b}(\frac{1}{-\partial^{2}})\varphi^{c} \nonumber \\
s'\omega^{a}&=&0   \hspace{1cm} (s')^{2}= 0.
\end{eqnarray}
The operator $s'$ exhibits explicit dependence from $v^{2}$
Moreover, it reduces to the operator $s$ when $v=0$. One should
notice that the operator $s'$ is non-local and representing a
symmetry of the action, that is an integrated functional, and cannot
be used to analyse the renormalizability properties of the model due
to his non-local character. However it can be very useful to
calculate the expectation value of exact BRST quantities as
presented in \cite{fewr}. It is also important to realize that
$\overline{\omega}^{a}\partial^{2}\omega^{a} -
\overline{\chi}^{a}\partial^{2}\chi^{a}$ is no longer cohomologicaly
trivial when we analyze the cohomology of $s'$. Thus terms that are
trivial by the $s$ symmetry becomes nontrial by $s'$ changing the
behavior of the propagator in the infrared limit.

Importantly, a Gribov type propagator has no direct interpretation
as particle. Thus the search for composed operators that presents a
particle representation, which are the natural candidates to be the
physical particles described in this action, are  fundamental in
order to obtain the physical content of the model. This new set of
symmetries despite being nonlocal are nilpotent and can formally
define a cohomological problem helping to obtain composite operators
that are good candidates to be associated to physical particles. For
example with this new symmetry is easy to observe that
\begin{equation}
\overline{\omega}^{a}{\omega}^{a} - \overline{\chi}^{a}{\chi}^{a}-
\frac{v^{2}}{\xi}\frac{1}{\partial^{2}}\varphi^{a}{\chi}^{a}
\end{equation}
has zero expectation value and is not a good candidate for a
composed operator. In fact is easy to use the property that this
term is a BRST variation of
\begin{equation}
s'(\overline{\omega}^{a}{\chi}^{a}) =
\overline{\omega}^{a}{\omega}^{a} - \overline{\chi}^{a}{\chi}^{a}-
\frac{v^{2}}{\xi}\frac{1}{\partial^{2}}\varphi^{a}{\chi}^{a},
\end{equation}
thus relating the expected value of
$\frac{v^{2}}{\xi}\frac{1}{\partial^{2}}\varphi^{a}{\chi}^{a}$ with
$\overline{\omega}^{a}{\omega}^{a} - \overline{\chi}^{a}{\chi}^{a}$.
This is precisely the property required to obtain a composite
operator which may have a particle representation. A good candidate
to have a particle representation should not be related to any other
composed operator by a BRST transformation. In fact this is a normal
property of the Slavnov-Taylor operator i.e variations of the
Slavnov-Taylor in relation to fields of the action turn possible to
obtain relations among correlators

Let us focus on dimension 2 condensates. According \cite{sps2} the
candidate to be a good composed operator is obtained imposing that
the correlator between those operators presents a spectral function
that is positive defined. In order to do that the
$K\ddot{a}ll\acute{e}n-Lehmann$ representation of the correlator
function is calculated for a scalar model. Our model reduces to that
presented in the above cited reference if we take the mass $m=0$,
$\frac{v^{2}}{\xi} = \mu^{2}$ and forget all interaction terms. In
these case the candidate to be a good composed operator to be
associated to a particle representation is given by
\begin{equation}
O= \varphi^{a}\varphi^{a} -
\frac{1}{\sqrt{2}}(\overline{\chi}^{a}-{\chi}^{a})(\overline{\chi}^{a}-{\chi}^{a}).
\end{equation}
This operator is not BRST invariant and his variation is given by:
\begin{equation}
s'O=-(\overline{\chi}^{a}-{\chi}^{a})\omega^{a} +
2\mu^{2}f_{abc}\theta^{b}(\overline{\chi}^{a}-{\chi}^{a})\frac{1}{\partial^{2}}\varphi^{c}.
\end{equation}
This operator is not a BRST variation but we are able to note that
\begin{eqnarray}
s'(O(x)O(y))&=&0 \nonumber \\
(O(x)O(y))&\neq & s'\Delta (x,y),
\end{eqnarray}
making the problem of obtaining a particle representation into a
cohomology problem.In order to turn more simple to understand this
result is useful to remember that the propagators of the toy model
are:
\begin{eqnarray}
<\varphi^{a}(k)\varphi^{b}(-k)>&=&\frac{k^{2}}{k^{4} + m^{2}k^{2} +
\mu^{4}} \hspace{1cm}<\varphi^{a}(k)\chi^{b}(-k)>=
\frac{\mu^{2}}{k^{4}+ m^{2}k^{2}+\mu^{4}}\nonumber \\
\hspace{1cm}<\varphi^{a}(k)\overline{\chi}^{b}(-k)>&=&
-\frac{\mu^{2}}{k^{4}+ m^{2}k^{2}+\mu^{4}}
\hspace{1cm}<\overline{\chi}^{a}(k)\chi^{b}(-k)>= \frac{k^{2} +
m^{2}}{k^{4} + m^{2}k^{2} +
\mu^{4}}\nonumber \\
<\overline{\omega}^{a}(k)\omega^{b}(-k)>&=&
-\frac{1}{k^{2}},\hspace{1cm} \mu^{2}=\frac{v^{2}}{\xi}.
\end{eqnarray}
The fact that does not exist a mixed propagator
$<\varphi^{a}\omega^{b}>$, due to the original quartet structure,
turn possible the existence of the above mentioned cohomology
problem for the physical correlator. This mechanism of understanding
the construction of physical correlators could be a way to see how
to generalize the $i-particle$ criterium presented in \cite{sps2} in
order to introduce interaction and go beyond the one loop
calculation.

\section{Conclusions}

\hspace{.7cm}In this work we present a model in which a nonlocal
term is introduced in a localized way and the one-loop effective
potential for the condensate
$<(\overline{\chi}^{a}-\chi^{a})\varphi^{b}>$ has been obtained by
combining the local composite operator (LCO), BRST quartets and an
algebraic renormalization. Our results indicate that in a model
constructed to obtain a Gribov type propagator, through an insertion
of a composite operator, the expectation value of this operator can
be obtained directly by the LCO technique and effective potential.
This result does not require geometric considerations in order to
obtain a gap equation as in the case of Gribov-Zwanziger and can be
applied to obtain the expectation value of any condensate of
ultraviolet dimension $2$ in $D=4$ actions. The spontaneous symmetry
breaking also opens a window of understanding for the determination
of observables whose definition makes sense only in the broken
phase.This is achieved through the BRS symmetry in the broken phase,
which although it can not be used for determining the
renormalizability of the model allows to correlate values expected
of operators already known in the unbroken phase with operators that
are not null only in the broken phase. This line of work in itself
opens the possibility to calculate expected values of several
composed operators that correspond to observables in a theory with
Gribov-type propagators.We hope to use this mechanism to obtain
observables in a gauge model in a future work.

\section{Acknowledgements}
The Conselho Nacional de Desenvolvimento Cient\'{\i}ico e tecnol\'{o}gico CNPq-
Brazil, Funda\c{c}\~{a}o de Amparo a Pesquisa do Estado do Rio de Janeiro
(Faperj) and the SR2-UERJ are acknowledged for the financial support.

\end{document}